\documentstyle[prl,aps,psfig]{revtex}
\begin{document}
\draft
\twocolumn[\hsize\textwidth\columnwidth\hsize\csname
@twocolumnfalse\endcsname

\widetext
\title{Resonating Valence Bond Wave Functions for Strongly Frustrated Spin Systems} 
\author{Luca Capriotti,$^{1}$ Federico Becca,$^{2}$  Alberto Parola,$^{3}$ and Sandro Sorella$^{4}$}
\address{${^1}$ Istituto Nazionale per la Fisica della Materia, Unit\`a di Firenze, I-50125 Firenze, Italy \\
${^2}$  Institut de Physique Th\'eorique, Universit\'e de Lausanne, CH-1015 Lausanne, Switzerland\\
${^3}$  Istituto Nazionale per la Fisica della Materia and Dipartimento di Scienze, Universit\`a dell'Insubria, I-22100 Como, Italy \\  
${^4}$  Istituto Nazionale per la Fisica della Materia, and SISSA, I-34014 Trieste, Italy\\ }
\date{\today}
\maketitle
\begin{abstract}
The Resonating Valence Bond (RVB) theory for two-dimensional 
quantum antiferromagnets is shown to be the correct paradigm for large 
enough  ``quantum frustration''. This scenario, proposed  long time 
ago but never confirmed by microscopic calculations, is very strongly 
supported by a new type of variational wave function, which is extremely 
close to the exact ground state of the $J_1{-}J_2$ Heisenberg model 
for $0.4 \lesssim J_2/J_1\lesssim 0.5$.
This wave function is proposed to represent the generic spin-half RVB ground state in   
spin liquids. 
\end{abstract}
]
\narrowtext

The question whether a frustrated spin-half system is well described by a 
spin-liquid ground state (GS) --  with no type of crystalline order --  
25 years after the first proposal \cite{fazekas} 
is still controversial, mainly because of the lack of reliable 
analytical or numerical solutions of model systems.
For unfrustrated or weakly frustrated quantum antiferromagnets 
a deep understanding of the nature of the GS together with 
a quantitative description of the ordered phase is obtained by 
including Gaussian quantum fluctuations over a classical N\'eel state. 
\cite{liu,franjo} 
For sizeable frustration, instead, this description is known to break down.
However, the short-range RVB state \cite{kivelson} does not prove a good starting point for the description of frustrated models;
it rather turns out to be the exact GS of {\em ad hoc} Hamiltonians.
\cite{kivelson,bose1,bose2}  

As a prototype of a realistic frustrated two-dimensional system,
which has been recently realized experimentally in $\rm Li_2VOSiO_4$ 
compounds,\cite{carretta}  we investigate the spin-half Heisenberg model 
with nearest ($J_1$) and next-nearest neighbor ($J_2$) superexchange 
couplings:
\begin{equation} \label{j1j2ham}
\hat{\cal{H}}=J{_1}\sum_{n.n.}
\hat{{\bf {S}}}_{i} \cdot \hat{{\bf {S}}}_{j}
+ J{_2}\sum_{n.n.n.}
\hat{{\bf {S}}}_{i} \cdot \hat{{\bf {S}}}_{j}~,
\end{equation}
on an $N-$site square lattice with periodic boundary conditions. 
In the ($J_2=0$) unfrustrated case, it is well 
established that the GS of the Heisenberg Hamiltonian 
has N\'eel long-range order, with a sizable value of the antiferromagnetic 
order parameter.\cite{calandra}
However, variational studies \cite{doucot} have shown 
that disordered, long-range RVB states have energies very close to the exact one.
It is therefore natural to imagine that by turning on the next-nearest 
neighbor interaction $J_2$, the combined effect of frustration 
and zero-point motion may eventually melt antiferromagnetism and stabilize a non-magnetic 
GS of purely quantum-mechanical nature.
Indeed, for $0.4 \lesssim J_2/J_1 \lesssim 0.6$ there is a general consensus
on the disappearance of the N\'eel order towards
a state whose nature is still much debated.\cite{russi} 

In a seminal paper,\cite{anderson} Anderson proposed 
that a physically transparent description of a RVB state 
can be obtained in fermionic representation
by starting from a BCS-type pairing wave function (WF), of the form
\begin{equation} \label{wavefunction}
|\psi_{\rm BCS}\rangle = {\rm exp} \left( \sum_{i,j}
f_{i,j} \hat{c}^{\dag}_{i,\uparrow} \hat{c}^{\dag}_{j,\downarrow} \right)
|0\rangle~.
\end{equation}                                 
This WF is the GS of the well-known BCS Hamiltonian
with a given (real) gap  function $\Delta_k=\Delta_{-k}$ provided
the Fourier transform $f_k$ of the pairing function, $f_{i,j}$, satisfies: 
$f_k = \Delta_k/(\epsilon_k + \sqrt{\epsilon_k^2 +\Delta_k^2})$,
where $\epsilon_k=-2 [\cos k_x + \cos k_y]$ is the free-electron dispersion. 
The non-trivial character of this WF emerges when we restrict 
to the subspace of fixed number of electrons (equal to the number of sites) and
enforce Gutzwiller projection onto the subspace with no double occupancies: 
singlet pairs do not overlap in real space and this 
WF can be described by a superposition of 
Valence Bond (VB) states.\cite{anderson,gros,poilblanc} 
Though this WF has been already studied  for the pure Heisenberg model by several authors\cite{gros,poilblanc}
for $\Delta_k \propto  (\cos k_x - \cos k_y)$, 
here we show that this type of RVB state  represents an  extremely 
accurate variational {\em ansatz} for {\em frustrated} systems.

A definite symmetry is guaranteed to the {\em projected} BCS ($p$-BCS)
state provided the gap function $\Delta_{k}$ 
transforms according to a one dimensional 
representation of the spatial symmetry group. A careful analysis
\cite{therevenge}, similar to the one carried out in\cite{poilblanc}, 
shows that the odd  component 
of the gap function $\Delta_{k}=-\Delta_{k+(\pi,\pi)}$
may have spatial symmetries different from 
those of the even component  $\Delta_{k}=\Delta_{k+(\pi,\pi)}$. 
We anticipate that the best variational energy is obtained when 
the former has $d_{x^2-y^2}$ symmetry, whereas the latter either vanishes or it
has $d_{xy}$ symmetry. In frustrated models, it is important 
to consider this generalization of the originally proposed 
WF,\cite{anderson,gros} because only in this way it is possible to
reproduce correctly the phases of the actual GS configurations.
In the unfrustrated case it is well known that such phases are
determined by the so-called Marshall-sign:
on each real space configuration $|x\rangle$,
the sign of the GS wave function
is determined only by the number of spin down in one of the two sublattices.    
This feature, rigorously valid for $J_2=0$, 
turns out to be a very robust property
for weak frustration ($J_2/J_1\lesssim 0.3$).\cite{richter} 
However, it is clearly violated when frustration plays an important role.
\begin{figure}
\centerline{\psfig{bbllx=65pt,bblly=225pt,bburx=500pt,bbury=620pt,%
figure=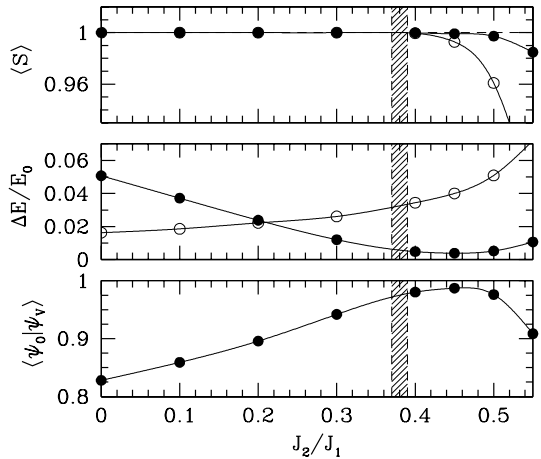,width=70mm,angle=0}}
\caption{\baselineskip .185in \label{over}
Average sign, accuracy of the GS energy, 
and overlap between the GS and the {\em p-}BCS state (full dots) as a function of
$J_2/J_1$, for $N=6\times 6$.
Empty dots are the Marshall sign (top panel), and the energy accuracy of
a N\'eel ordered spin-wave WF \protect\cite{franjo}
(middle panel). Lines are guides for the eye and the shaded region indicates  
the location of the expected transition point  to the non-magnetic phase.
}
\end{figure}
In order to determine the best variational WF of
this form we have used a recently developed quantum Monte Carlo (QMC) 
technique \cite{flst} that allows to optimize a large number
of variational parameters with modest computational effort.
We first consider the largest square cluster $N=6\times6$ where the exact 
GS can be numerically determined by exact diagonalization (ED).
In order to show the quality of the present WF  when 
frustration ($J_2/J_1$) is increased, we have computed the variational 
energy, the overlap of the $p$-BCS wave function with the exact GS, $|\psi_0\rangle$, 
and the average sign, defined for a generic variational state  $|\psi_V\rangle$ as 
$\langle S\rangle =  \sum_x |\langle x|\psi_{\rm V}\rangle|^2
{\rm Sgn} \big[ \langle x|\psi_{\rm V}\rangle \langle x|\psi_0\rangle \big]~.
$
The Marshall sign (i.e., $\langle S \rangle =1$ for $J_2=0$) 
is obtained using the  
$p$-BCS wave function, with only the $d_{x^2-y^2}$  component.
However, for $J_2/J_1 \gtrsim 0.4$, the phases of the 
WF are considerably affected by the strong frustration 
and only when a 
sizable $d_{xy}$ component is stabilized at the variational level, 
this property 
can be correctly reproduced, as clearly shown in Fig.~\ref{over}.
Remarkably, as shown in the same figure,
this kind of WF is not only 
in qualitative agreement with the exact solution but 
it is also impressively accurate in the region $J_2/J_1 \sim 0.45 \pm 0.05$ 
of large frustration, where the overlap of the variational WF 
is improved by more than an order of magnitude 
with respect to the $J_2=0$ case.
This fact implies that the GS in the strongly 
frustrated regime is almost exactly reproduced by a RVB wave function, 
at least on clusters of this size.

\begin{figure}
\centerline{\psfig{bbllx=110pt,bblly=250pt,bburx=475pt,bbury=525pt,%
figure=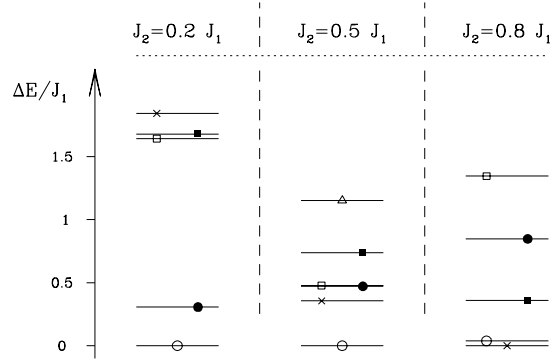,width=70mm,angle=0}}
\caption{\baselineskip .185in \label{levels}
Lowest energy states with given quantum numbers
referenced to the GS energy:
Singlet zero momentum $s$-wave (empty dot)  and $d$-wave (cross),
singlet at momentum $(0,\pi)$ (empty square),
singlet at momentum $(\pi,\pi)$ (empty triangle),
triplets at momentum $(\pi,\pi)$ $s$-wave (full dot)
and at momentum $(0,\pi)$ (full square).
}
\end{figure}

Further indications of the changes in the nature of the GS 
occurring by increasing the frustration ratio 
can be found in the ordering of
states with different quantum numbers in the
energy spectrum.\cite{bernu}
This information can be easily
accessed by ED, which has
been performed on the $6\times 6$ cluster for 
three representative values of the frustration ratio:
$J_2/J_1=0.2; 0.5; 0.8$.
In the case of N\'eel order the two lowest states of the 
finite-size spectrum are a 
total-symmetric singlet and a triplet of 
momentum $(\pi,\pi)$. This phase is clearly expected to occur 
for sufficiently small $J_2/J_1$. Analogously, in 
the large $J_2/J_1$ limit, the two sublattices decouple
and a collinear state characterized by ferromagnetic
stripes, staggered along the direction orthogonal to the stripe, 
is believed to prevail.\cite{chandra} In this case, the symmetry
breaking implies that four classes of states with
different spatial symmetries become degenerate:
the lowest representatives of these families are
an $s$-wave and a $d$-wave singlet at zero momentum
and two triplets at momenta $(0,\pi)$ and $(\pi,0)$.
Therefore the transition between these two 
ordered phases implies that, by increasing $J_2/J_1$,
the $(\pi,\pi)$ triplet should acquire a
gap while the $d$-wave singlet and the $(0,\pi)$ and $(\pi,0)$
triplets should collapse onto the GS.
The low-energy spectrum is shown in   Fig.~\ref{levels},
suggesting that the reshuffling of the lowest energy levels in the system 
occurs at least in two steps: first the triplet levels lift,
leaving room for a non-magnetic phase with a finite triplet gap
and then the $d$-wave singlet collapses.

In principle either a homogeneous spin liquid or a VB crystal 
with some broken spatial symmetry is compatible with a triplet gap in the
excitation spectrum. Indeed, there are several proposals 
that the GS for $J_2/J_1\sim 0.5$ may be 
spontaneously dimerized \cite{russi,singh,read} with broken 
translation-rotation symmetry (columnar VB state) as it
is likely to happen in a generalization
of the present model.\cite{bose2} Alternatively,
a plaquette VB state \cite{zhitomirsky,luca} 
 with broken translational symmetry but preserving
rotational symmetry may be stabilized.
Note that a plaquette VB state would imply the degeneracy
of four singlet states at momenta $(0,0)$, $(\pi,0)$, $(0,\pi)$ 
and ($\pi,\pi)$, while a columnar VB state \cite{russi,singh,read} would result 
from the mixing of four singlets with different quantum numbers:
two translationally invariant $s$-wave and  $d$-wave states and 
the two singlets at momenta $(\pi,0)$ and $(0,\pi)$.
The plaquette scenario recently proposed\cite{zhitomirsky,luca} 
turns out to be very unlikely in this model because 
the lowest singlet of momentum $(\pi,\pi)$ lies very high in energy 
at all the couplings we have investigated.
The presence of a d-wave singlet in the singlet-triplet gap, instead,
has been also evidenced in the regime of strong frustration 
of the $J_1{-}J_2$ model on the 1/5-depleted square lattice, 
where the ground state 
is believed to be spin liquid.\cite{albrecht}
On the other hand on the basis of these ED data,
it is clearly impossible to  establish whether the GS is  
dimerized or disordered. 
\begin{figure}
\centerline{\psfig{bbllx=60pt,bblly=160pt,bburx=520pt,bbury=715pt,%
figure=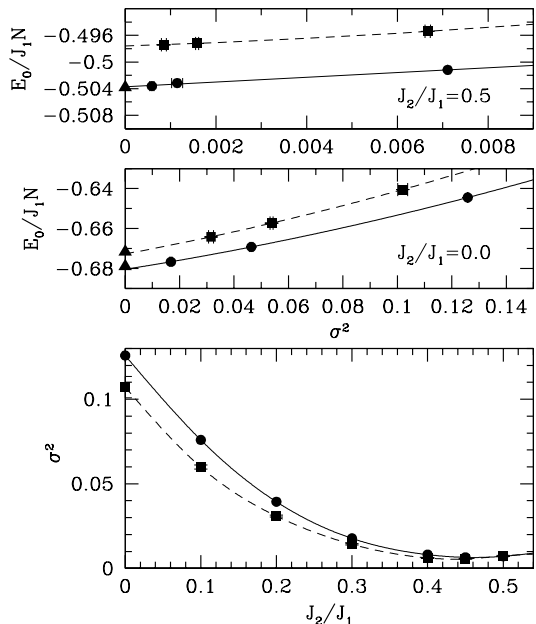,width=70mm,angle=0}}
\caption{\baselineskip .185in \label{variance}
Upper panels:  energy vs. decreasing  variance for  
the $p$-BCS wave function with zero, one and two Lanczos   
iterations. ED and QMC \protect\cite{calandra} (full triangles), 
are shown for comparison. 
Lower panel: variance vs. $J_2/J_1$. 
In all the plots: $N=6\times 6$ (dots) and, $N=10\times 10$ (squares).}
\end{figure}
In order to clarify  this issue we have extended the 
calculation to much larger system size, by also 
employing a few Lanczos iterations
over the starting variational WF.
The stochastically implemented Lanczos technique 
is a new QMC method with very good convergence properties
when the initial WF well represents the actual GS.\cite{flst} 
This accuracy can be confirmed {\em a priori} even on large size, 
by studying the variance $\sigma^2= 
(\langle \hat{\cal{H}}^2\rangle -\langle\hat{\cal{H}}\rangle^2)/J_1^2N$ of the energy, 
the variance being smaller (zero) for a better (exact) calculation.
As shown in Fig.~\ref{variance}, also in the 
$10\times 10$ cluster the variance as a function of $J_2/J_1$ 
behaves similarly to the $6\times 6$ case, 
strongly suggesting that the exceptional accuracy 
of the $p$-BCS wave function does not decrease
for larger sizes. This is also confirmed by
the extremely well-behaved approach to the zero-variance  
limit with few Lanczos iterations (shown in the same figure),
leading to an almost exact estimate of the GS energy even for 
$J_2=0$, when the accuracy is the lowest.
\begin{figure}
\centerline{\psfig{bbllx=40pt,bblly=260pt,bburx=500pt,bbury=645pt,%
figure=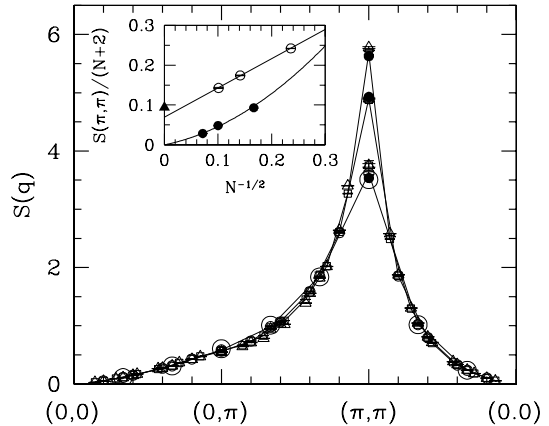,width=70mm,angle=0}}
\caption{\baselineskip .185in \label{sq}
$S(q)$ for  (from the lower to the upper curve) $N=6\times 6$, $10\times 10$, and $14\times 14$. 
Variational estimate (empty triangles), 
with one Lanczos iteration (empty squares), with two Lanczos iterations (empty dots). 
Full dots: variance-extrapolated values of $S(\pi,\pi)$; large empty circles: ED results. 
Inset: size-scaling of the variance-extrapolated order parameter 
squared for $J_2/J_1=0.5$ (full dots) and $J_2=0$ (empty dots, Ref.~\protect\cite{flst}). 
The full triangle is the thermodynamic value for $J_2=0$ 
taken from Ref.~\protect\cite{calandra}.
}
\end{figure}
Of course, the accuracy in the energy does not necessarily 
guarantee a corresponding accuracy in correlation functions. 
However, in the strongly frustrated regime, our  approach is particularly reliable 
since the gap to the first excitation belonging to the 
same subspace of our best WF (with two Lanczos iterations) is bounded 
in all the most plausible cases (columnar, plaquette, non-degenerate 
singlet RVB) by the triplet gap 
($\gtrsim 0.1J_1$,\cite{russi,luca}) and therefore
is much larger than the estimated error in the total energy ($\sim 0.01J_1$, 
see top panel of Fig.~\ref{variance}). 
Indeed, as shown in Fig.~\ref{sq}, the comparison 
of the  magnetic structure factor
$S(q)=\langle \hat{{\bf{S}}}_{q} \cdot \hat{{\bf{S}}}_{-q} \rangle$, 
with the exact result gives
a clear indication that correlation functions obtained by the variational 
approach are essentially exact, indicating also the absence of 
long-range N\'eel order.\cite{gros} This fact is particularly evident 
because for all the lattice sizes considered,  $S(\pi,\pi)$ 
is slightly depressed by few Lanczos iterations, meaning that 
the exact value of the magnetic structure factor  is bounded by the one of 
the $p$-BCS wave function, with sizable antiferromagnetic correlations 
but with no antiferromagnetic long-range order.
Remarkably, correlation functions are smoothly depending on 
the energy variance, so that an estimate of the magnetic order parameter 
within $\sim 10\%$ can be achieved also 
for $J_2=0$, where our singlet WF  is not
particularly accurate and the spectrum is gapless 
(see the inset of Fig.~\ref{sq}).

\begin{figure}
\centerline{\psfig{bbllx=30pt,bblly=130pt,bburx=570pt,bbury=695pt,%
figure=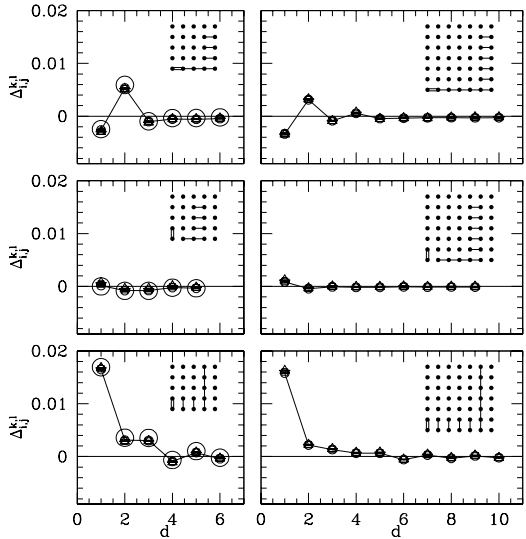,width=70mm,angle=0}}
\caption{\baselineskip .185in \label{dimer}
Dimer-dimer correlation functions $\Delta_{i,j}^{k,l}$ obtained by keeping
fixed the position of the left-most bond $(i,j)$ (double stick) and
moving the bond $(k,l)$ (single stick) along the indicated patterns. 
$d$ is the Manhattan distance. $6\times 6$ (left), $10\times 10$ (right);
symbols as in Fig.~\ref{sq}. 
}
\end{figure}

In order to investigate the existence of long-range dimer-like correlations,
as in the columnar 
or the plaquette VB state, we have calculated
the dimer-dimer correlation functions, 
$\Delta_{i,j}^{k,l}=\langle \hat{S}_i^z\hat{S}_j^z \hat{S}_k^z \hat{S}_l^z\rangle - \langle \hat{S}_i^z\hat{S}_j^z \rangle \langle \hat{S}_k^z \hat{S}_l^z\rangle$. 
In presence of a broken spatial symmetry, the latter  should converge 
to a finite value for large distance. 
This is clearly ruled out by our results, shown 
in Fig.~\ref{dimer}, with a very robust indication of the 
{\em liquid}  character of the GS for $J_2/J_1\simeq 0.5$, which is correctly 
described by our variational approach.

A totally symmetric spin-liquid solution proposed for this 
model in Ref.~\cite{kivelson} was actually rather unexpected 
after the work of Read 
and Sachdev,\cite{read} providing arguments in favor of spontaneous 
dimerization. This conclusion was supported by series expansion \cite{russi,singh}
and QMC studies included the one done  by two of us.\cite{luca} 
It is clear however that it is very hard 
to reproduce a fully symmetric spin liquid GS, with  any technique, numerical 
or analytical, based on reference states explicitly breaking 
some lattice symmetry.\cite{note}
We do not know whether numerical methods and/or series expansions 
can ever solve this controversial issue. 
However, at the time being, we can safely state that in order 
to have something different from a spin-liquid GS, one has 
necessarily to improve the quality of our spin-liquid variational 
WF, e.g., with small symmetry breaking terms; a possibility 
that we have attempted (small dimerizations  or 
plaquette-like perturbations) without success. 
Indeed, this seems a very difficult task due to
the tiny energy range left 
($\sim 10^{-3}J_1$ per site) by our variational WF. 

In conclusion, the spin-liquid RVB ground state, originally proposed to 
explain high-Temperature superconductivity, is indeed very plausible
for strongly frustrated spin systems.  
We expect that the $p$-BCS Resonating Valence Bond wave function 
represents the  {\em generic}  variational state
for a spin-half spin liquid, once the pairing function $f_{i,j}$ 
is exhaustively parameterized
according to the symmetries of the Hamiltonian.
In particular, the $p$-BCS wave function
can be easily extended 
to the case of topologically frustrated lattices -- like
the kagom\'e or the pyrochlore lattices -- as
well as to frustrated models on the 1/5-depleted square lattice.

We thank F. Mila and C. Lhuillier for useful discussions;  
one of us (S.S.) acknowledges
the ETH-Z\"urich for the kind hospitality.
This work has been partially supported by MURST (COFIN99).



\begin{thebibliography}{99}

\bibitem{fazekas}  P. Fazekas and P.W. Anderson, Philos. Mag. {\bf 30}, 423 (1974).

\bibitem{liu} Z. Liu and E. Manousakis, \prb {\bf 40}, 11437 (1989).

\bibitem{franjo} F. Franjic and S. Sorella, Prog. Theor. Phys. {\bf 97}, 399 (1997).

\bibitem{kivelson} F. Figueirido {\em et al.}, \prb {\bf 41}, 4619 (1989).

\bibitem{bose1} I. Bose and A. Ghosh, \prb {\bf 56}, 3149 (1997).

\bibitem{bose2} I. Bose and P. Mitra, \prb {\bf 44} 443 (1991).


\bibitem{carretta} R. Melzi {\em et al.}, \prl {\bf 85}, 1318 (2000).  

\bibitem{calandra} A.W. Sandvik, \prb {\bf 56}, 11678 (1997). 

\bibitem{doucot} S. Liang {\em et al.}, \prl {\bf 61}, 365 (1988).

\bibitem{russi}  V.N. Kotov {\em et al.}, Phil. Mag. B {\bf 80}, 1483 (2000).

\bibitem{anderson} P.W. Anderson, Science {\bf 235}, 1196 (1987). 

\bibitem{gros} C. Gros, \prb {\bf 38}, 931 (1988). 
 
\bibitem{poilblanc} D. Poilblanc, \prb {\bf 39}, 140 (1989). 

\bibitem{therevenge} L. Capriotti {\em et al.}, in preparation.

\bibitem{richter} J. Richter {\em et al.}, Europhys. Lett. {\bf 25}, 545 (1994).

\bibitem{flst} S. Sorella,  \prb (in press).

\bibitem{bernu}  B. Bernu {\em et al.}, \prl {\bf 69}, 2590 (1992).

\bibitem{chandra} P. Chandra {\em et al.}, \prl {\bf 64}, 88 (1990).

\bibitem{singh} R.R.P. Singh {\em et al.}, \prb {\bf 60}, 7278 (1999).

\bibitem{read} N. Read and S. Sachdev, \prl {\bf 66}, 1773 (1991).

\bibitem{zhitomirsky} M. Zhitomirsky and K. Ueda, \prb {\bf 54}, 9007 (1996).

\bibitem{luca} L. Capriotti and S. Sorella, \prl {\bf 84}, 3173 (2000). 

\bibitem{albrecht} M. Albrecht {\em et al.}, \prb {\bf 54}, 15856 (1996).

\bibitem{note} 
In Ref.~\cite{luca}, the plaquette scenario was suggested by 
the behavior of the susceptibility, calculated in presence of an external 
field $h$ coupled to the dimer operator. However, recent ED calculations 
have shown that the $h\to 0$ limit present some subtleties that have lead 
to an overestimate of the susceptibility.\cite{therevenge}

\end{thebibliography}
\end{document}